\documentclass[journal]{IEEEtran}

\ifCLASSINFOpdf

\else

\fi

\ifCLASSINFOpdf
\usepackage[pdftex]{graphicx}
\graphicspath{{../pdf/}{../jpeg/}}
\DeclareGraphicsExtensions{.pdf,.jpeg,.png}
\else

\usepackage[dvips]{graphicx}\graphicspath{{../eps/}}
\DeclareGraphicsExtensions{.eps}
\fi

\usepackage{amsmath}
\usepackage{algorithmic}
\usepackage{array}
\ifCLASSOPTIONcompsoc
 \usepackage[caption=false,font=footnotesize,labelfont=sf,textfont=sf]{subfig}
\else
\usepackage[caption=false,font=footnotesize]{subfig}
\fi

\usepackage{fixltx2e}
\usepackage{dblfloatfix}
\usepackage{url}
\usepackage[table,xcdraw]{xcolor}
\usepackage{multirow}
\usepackage{tabularx}
\usepackage{longtable}
\usepackage{tabularx}
\usepackage{pifont} 
\usepackage{float}
\newcolumntype{Y}{>{\centering\arraybackslash}X}

\begin{document}
%
\title{Internet of Things Security, Device Authentication and Access Control: A Review}
%
%
%

\author{Inayat~Ali,
        Sonia~Sabir,
        and~Zahid~Ullah,
\thanks{Inayat Ali and Sonia Sabir were with the Department of Computer Science at the COMSATS University of Science and Technology, Islamabad, Pakistan}
\thanks{Zahid Ullah was with the Department of Electrical Engineering at the COMSATS University of Science and Technology, Islamabad, Pakistan}
\thanks{\textit{For correspondance: falcon19khan@gmail.com} \\Manuscript received July 23, 2016; Published August  2016.}}


\maketitle

\begin{abstract}
The Internet of Things (IoT) is an emerging technology that has garnered significant attention from both academia and industry. IoT involves the interconnection of internet-enabled devices, allowing them to communicate with each other and with humans to achieve common objectives. In the near future, IoT is expected to be seamlessly integrated into our daily lives, making humans increasingly reliant on this technology for convenience and an improved lifestyle. Any security breach within IoT systems could directly impact human well-being, making security and privacy critical concerns that must be addressed.

In this paper, we present a comprehensive study of security challenges in IoT and classify potential cyberattacks across different layers of the IoT architecture. We also explore the limitations of traditional security solutions, such as cryptography, authentication mechanisms, and key management, within the context of IoT. One crucial yet under-explored aspect of IoT security is device authentication and access control. In this work, we consolidate the latest advancements in device authentication and access control techniques, providing a thorough overview in a single document.
\end{abstract}

\begin{IEEEkeywords}
Internet of Things, Authentication, Access Control, Security, Cyber-attacks, Wireless Sensor Networks

\end{IEEEkeywords}
\IEEEpeerreviewmaketitle

\section{Introduction}
The Internet of Things (IoT) is an emerging technology that focuses on the interconnection of devices and objects with each other and with humans to achieve common goals. IoT is powered by various existing technologies such as Wireless Sensor and Actuator Networks (WSAN) and Radio Frequency Identification (RFID). The concept of IoT was first introduced by Kevin Ashton at the Auto-ID Center at MIT \cite{ref1}. With the widespread availability of the internet through Wi-Fi and mobile data networks (3G, 4G LTE), ubiquitous sensing has become a reality, paving the way for enhanced connectivity among devices and users, which is expected to contribute significantly to the development of smart cities in the future. By 2020 \cite{ref2}, the number of connected devices is projected to reach between 50 and 100 billion, leading to ubiquitous sensing and a wide range of services.

In the IoT paradigm, information and communication systems will be seamlessly embedded in our environment, enabling the sensing and processing of various physical phenomena and storing the resulting data on remote clouds \cite{ref3}. IoT is essential for developing smart homes, cities, and healthcare systems. However, widespread adoption of IoT will depend on the technology's ability to gain users' trust by ensuring robust security and privacy.

IoT security is currently a hot research topic, with researchers worldwide addressing the various challenges posed by the heterogeneous nature of IoT. Since IoT integrates multiple technologies, each with its traditional security and privacy vulnerabilities, these issues must be addressed in the IoT context. This paper provides a brief overview of the IoT architecture for assessing security at each layer and discusses the security threats and potential attacks that adversaries could launch. We also propose countermeasures to mitigate these risks.

IoT infrastructure is vulnerable to well-known security attacks such as Denial of Service (DoS), replay attacks, man-in-the-middle attacks, device cloning, eavesdropping, and routing attacks \cite{ref4}. Specific IoT-related cyberattacks, such as device tampering, privacy breaches, information disclosure, DoS, spoofing, signal injection, and side-channel attacks, have also been identified \cite{ref5}. IoT devices are typically resource-constrained, making it challenging to implement cryptographic security solutions, which in turn exposes them to data integrity and confidentiality issues. Additionally, the large number of devices and the machine-to-machine (M2M) communication nature of IoT present unique challenges to traditional security solutions, including device authentication and access control mechanisms. This paper discusses these challenges in detail and briefly reviews recently proposed techniques for device authentication and access control.

The IoT has a broad range of applications, including smart homes, cities, healthcare systems, intelligent traffic control, connected vehicles, environmental monitoring, smart grids, metering, water network monitoring, and smart logistics, among others \cite{ref3} \cite{ref5}. While the application scope of IoT extends beyond these examples, this paper focuses on generic security issues applicable across all IoT domains.

\subsection{Motivation}
 For the IoT to be widely adopted by the industry, it must earn the trust of users by ensuring robust security and privacy measures. Although IoT security is an active area of research, there is a scarcity of up-to-date comprehensive reviews on the subject, with existing studies often outdated and not reflecting the latest threats that frequently emerge in the IoT landscape \cite{ref18}\cite{ref40}. This gap highlights the need for a current and thorough review of IoT security to guide researchers in focusing their efforts on specific security challenges. Additionally, support layer security in IoT has not been adequately addressed in existing literature. Our work aims to fill this gap by identifying and discussing various support layer security issues. Furthermore, we provide an in-depth study of the latest developments in authentication and access control mechanisms, which remain critical challenges in the IoT domain.

 The rest of the paper is organized as follows: Section 2 discusses the IoT architecture. In Section 3, we present a comprehensive study of security challenges in IoT. Section 4 addresses the challenges posed to traditional security solutions in the context of IoT. In Section 5, we explore the state-of-the-art authentication and access control mechanisms used in the IoT. Finally, we conclude our work in Section 6.

\section{INTERNET OF THINGS ARCHITECTURE}

The IoT is poised to reshape the world, offering significant enhancements to human life. However, ensuring the security of IoT is both crucial and challenging due to its heterogeneous nature, extensive deployment, resource-constrained nodes, and the generation of vast amounts of data every second. The IoT network architecture typically comprises four layers \cite{ref14}, as illustrated in Fig. \ref{fig1}. Although this is not a standardized architecture, most proposed IoT architectures include these layers, making it a useful reference for identifying and classifying various security issues within IoT. The architecture depicted in Fig. \ref{fig1} represents the most widely accepted structure, with the following layers:

\begin{figure}[!t]
\centering
\includegraphics[width=2.6in]{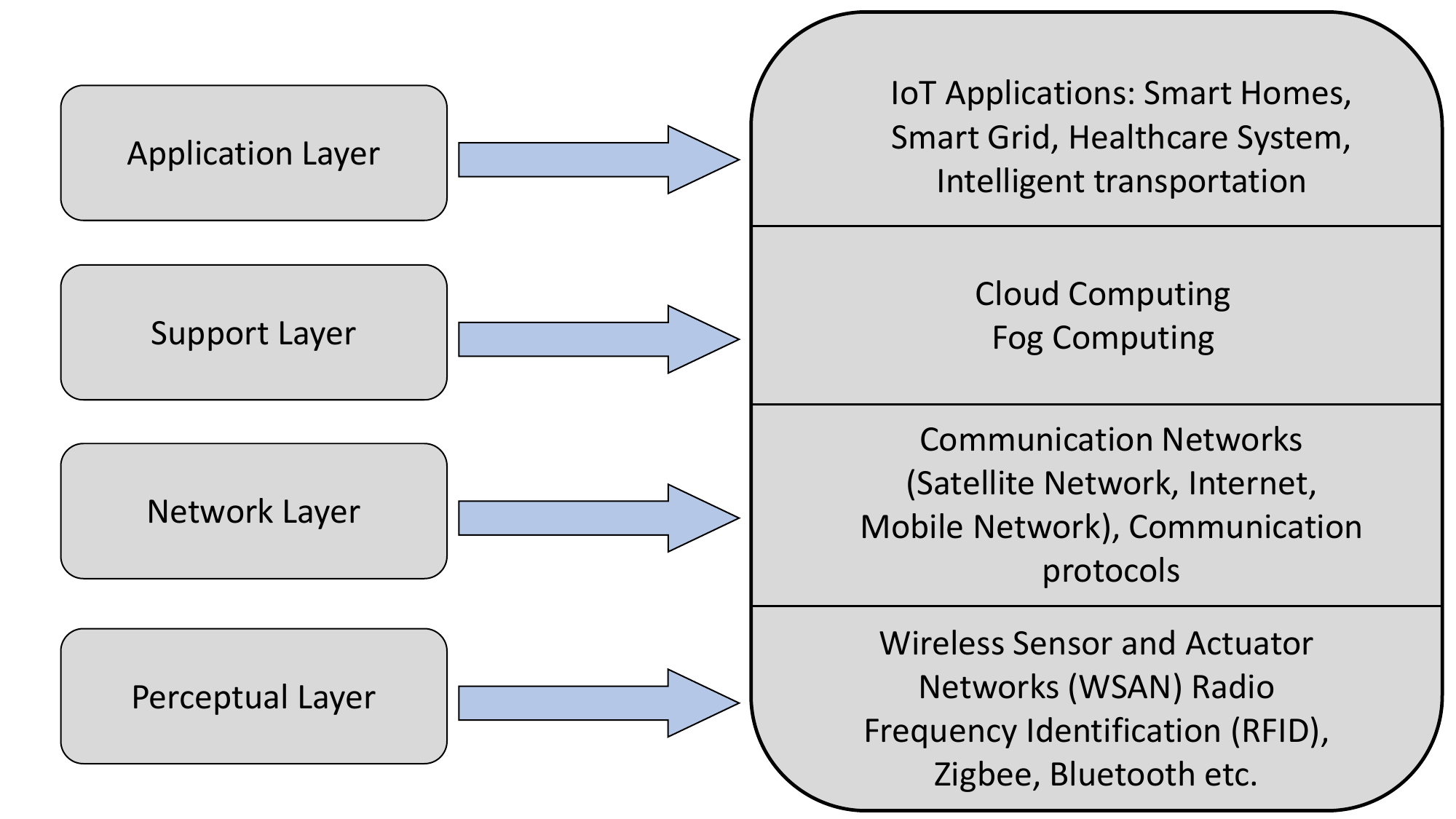}
\caption{IoT architecture}
\label{fig1}
\end{figure}

\subsection{Perceptual Layer }
This layer comprises devices such as sensors and RFID tags that detect and monitor real-world physical phenomena, such as RFID-tagged objects, weather conditions, and water levels in agricultural fields. Wireless Sensor and Actuator Networks (WSAN) and Radio Frequency Identification (RFID) are the key components of this layer, enabling the collection and transmission of data critical to IoT operations. These devices play a fundamental role in gathering the data that drives IoT systems, providing the essential input needed for subsequent processing and decision-making in the higher layers of the architecture.

\subsection{Network Layer}
This layer is responsible for securely transmitting the data collected by devices in the perceptual layer to fog nodes, the main cloud, or directly to another IoT node. It utilizes various technologies such as mobile networks, satellite networks, and Wireless Ad Hoc Networks, along with a range of secure communication protocols employed in these technologies. The secure transmission of data is critical in maintaining the integrity and confidentiality of information as it moves across the IoT ecosystem, ensuring that the data reaches its destination reliably and securely.

\subsection{Support Layer}
The support layer provides a robust and efficient platform for IoT applications. It enables various IoT applications to be hosted on fog nodes or the main cloud, making them accessible to resource-constrained devices via the internet. This layer offers essential storage and computing power to these devices, compensating for their limited resources. By offloading data processing and storage to the cloud or fog nodes, the support layer ensures that even devices with minimal computational capabilities can participate in the IoT ecosystem, enhancing their functionality and the overall system's scalability.

\subsection{Application Layer}

This layer delivers IoT services to users based on their specific needs. Users can access various services through the application layer interface, which serves as the gateway to different IoT applications. These applications include smart homes, smart healthcare systems, intelligent transportation, smart agriculture, automated vehicles, and many more. The application layer tailors services to individual user requirements, ensuring that IoT solutions are practical, user-friendly, and effective in improving daily life across a wide range of domains.

\section{SECURITY IN THE INTERNET OF THINGS}

\begin{figure*}[!t]
\centering
\includegraphics[width=6.6in]{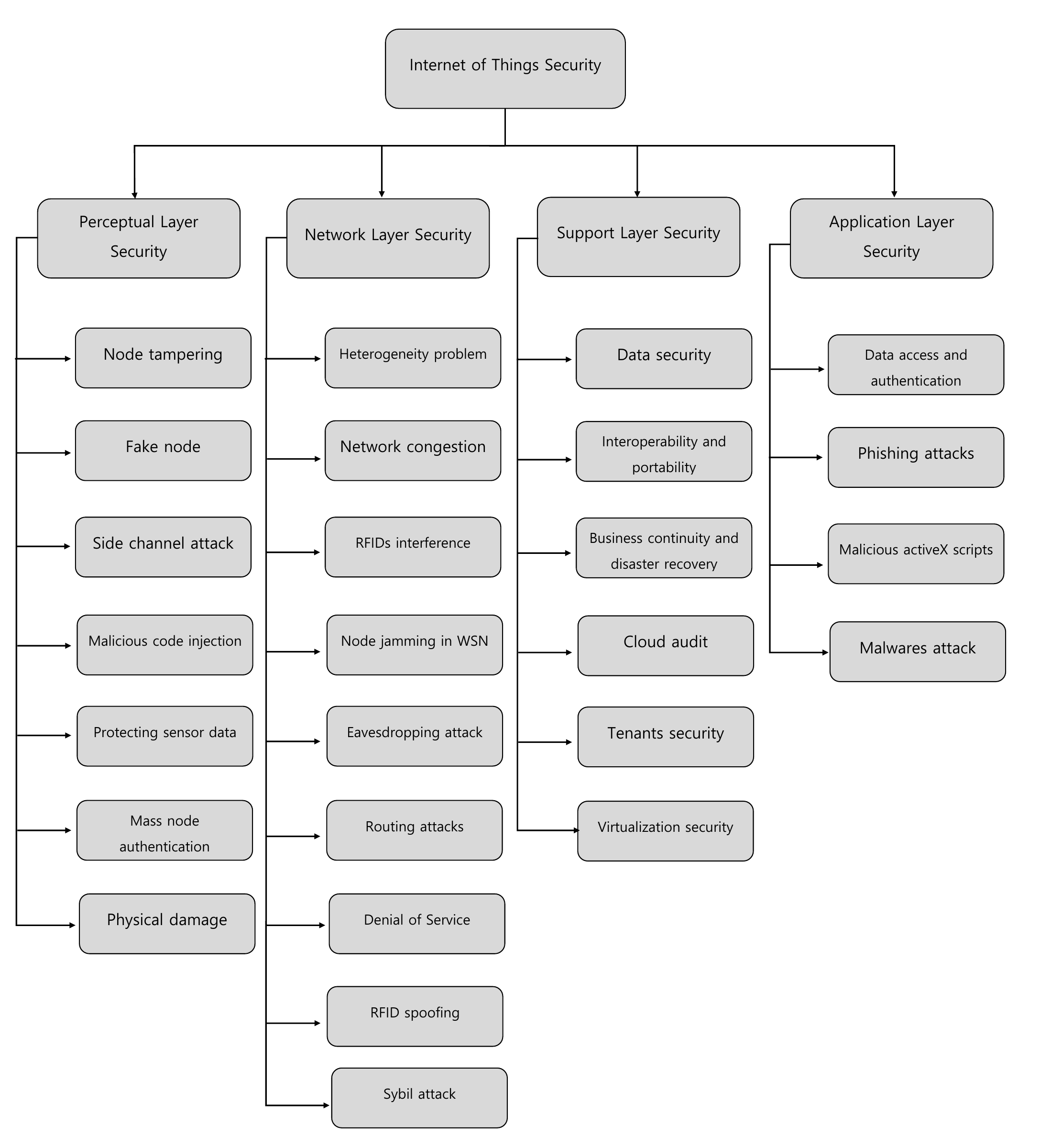}
\caption{IoT security and attacks}
\label{fig2}
\end{figure*}

Despite the significant importance and broad applications of IoT, deploying it in mission-critical areas poses substantial challenges, particularly where security and privacy are paramount concerns. For instance, a successful security attack on a smart healthcare system could result in the loss of numerous patient lives, while an attack on an intelligent transportation system could lead to severe financial losses and fatalities. Ensuring the security of IoT in such critical applications is a complex issue that requires further research to address these challenges effectively. In this paper, we discuss these security challenges in the context of IoT architecture. Fig. \ref{fig2} provides a visual summary of these security issues, highlighting the key areas of concern.

\subsection{Perceptual layer security}
The perceptual layer is composed of resource-constrained IoT devices such as sensors, RFID tags, Bluetooth, and Zigbee devices. These devices are particularly vulnerable to cyber-attacks due to their limited processing power and memory. Additionally, because many IoT devices are physically deployed in open or unsecured environments, they are susceptible to a variety of physical attacks. Some of the common physical attacks these devices may encounter include:

\subsubsection{Node tempering}

If an attacker gains physical access to sensor nodes, they can potentially replace the entire node or parts of its hardware, or directly connect to it to alter sensitive information and gain unauthorized access to the node [15]. This sensitive information could include cryptographic keys or routes in the routing table, which are crucial for maintaining the security and integrity of the IoT network. Such breaches can severely compromise the network's security and make it vulnerable to further attacks.

\subsubsection{Fake node}
An attacker can introduce a fake node into the IoT system and inject malicious data through this fake node, causing low-power devices to become overly busy and deplete their energy resources [18]. This type of attack can also be leveraged as a man-in-the-middle attack, further compromising the security and integrity of the IoT network.

\subsubsection{Side channel attack}
Attackers can exploit information such as power consumption, timing, and electromagnetic radiation emitted by sensor nodes to compromise encryption mechanisms [18]. These side-channel attacks allow attackers to infer sensitive data, weakening the overall security of the IoT system.

\subsubsection{Physical damage}
An adversary can physically damage IoT devices to launch a DoS attack. Since IoT devices are deployed in both open and enclosed environments, they are particularly vulnerable to physical damage by attackers, making them susceptible to such attacks. Physical tampering can render the devices inoperable, effectively disrupting the network and its services.

\subsubsection{Malicious code injection}
An adversary can physically compromise a node by inserting malicious code, granting them unauthorized access to the system [41]. This type of attack can undermine the security of the entire IoT network, allowing the attacker to manipulate or disrupt network operations.

\subsubsection{Protecting sensor data}
The confidentiality requirements of sensor data in an IoT system are relatively low, as an adversary can simply place another sensor nearby to obtain the same readings. However, ensuring the integrity and authenticity of this data is critically important and must be secured. This is because any compromise in the integrity or authenticity of the data could lead to incorrect decisions or actions based on falsified information, which could have significant consequences for the system's operation and security.

\subsubsection{Mass node authentication}
A large number of nodes in an IoT system experience authentication challenges [18]. The substantial amount of network communication required solely for authentication purposes can negatively impact the system's overall performance, leading to increased latency and resource consumption. This highlights the need for efficient and scalable authentication mechanisms to maintain the performance and security of IoT networks.

\subsubsection{Security requirements of perceptual Layer}

To ensure the security of an IoT system, it is crucial to first physically secure the system against unauthorized access by adversaries. Node authentication is also essential to prevent illegal access to the system, ensuring that only authorized devices can communicate within the network. Additionally, the integrity and confidentiality of data transmitted between nodes are vital, necessitating the design of lightweight cryptographic algorithms tailored for secure data transmission in resource-constrained environments like IoT. Key management remains a significant challenge that must be addressed to ensure secure and efficient operation within IoT networks.

\subsection{Network Layer Security}
While the core network of IoT systems is equipped with robust security measures, certain vulnerabilities still persist. Traditional security issues can compromise the integrity and confidentiality of data. Various network attacks, such as eavesdropping, DoS, Man-in-the-Middle attacks, and virus invasions, continue to pose significant threats to the network layer. These attacks can disrupt communication, steal sensitive information, and undermine the overall security of the IoT network, highlighting the need for ongoing vigilance and the development of advanced security strategies to protect against these threats.

\subsubsection{Heterogeneity problem}

The IoT perceptual layer is composed of a wide range of heterogeneous technologies. The access network employs multiple access methods, and this heterogeneity significantly increases the challenges related to security and interoperability [18]. This diversity in technology and communication protocols makes it difficult to create uniform security measures across all devices and systems, leading to potential vulnerabilities that adversaries could exploit. Addressing these challenges requires developing security solutions that are both flexible and capable of operating effectively across diverse technologies within the IoT ecosystem.

\subsubsection{Network congestion problems}

The large volume of sensor data, combined with the communication overhead caused by the authentication of numerous devices, can lead to network congestion [18]. To address this issue, it is essential to implement an efficient device authentication mechanism and adopt robust transport protocols capable of managing the high data load. These measures will help to alleviate congestion and ensure smooth and secure communication within the IoT network.

\subsubsection{RFIDs interference}
This is a network layer attack where the radio frequency signals used by RFID devices are disrupted by noise signals, leading to a DoS [16]. This interference prevents the RFID devices from communicating effectively, causing a breakdown in the system's functionality and rendering the affected devices inoperative.

\subsubsection{Node jamming in WSN}
This attack is similar to the radio frequency interference described above for RFID systems. In this type of attack, the attacker disrupts the radio frequency signals used by wireless sensor networks (WSNs), leading to a DoS [17]. By interfering with the radio frequencies, the attacker effectively prevents the WSNs from functioning properly, thereby denying services and causing significant disruption to the network's operations.

\subsubsection{Eavesdropping attack}
This attack involves the sniffing of traffic within the wireless vicinity of Wireless Sensor Networks (WSNs), RFIDs, or Bluetooth devices due to the inherently wireless nature of the device layer in IoT [19]. Such attacks often begin with information gathering through sniffing, using tools like packet sniffers [20]. By intercepting and analyzing the wireless communications, attackers can gather sensitive information, which can then be used to launch further attacks against the IoT network.

\subsubsection{Denial of service}
The adversary overwhelms the network with traffic exceeding its capacity, rendering the network unavailable for legitimate users and disrupting essential services. This type of attack, often referred to as a DoS attack, effectively prevents legitimate traffic from accessing network resources by flooding the network with excessive, illegitimate requests or data.

\subsubsection{RFID spoofing}
The attacker spoofs RFID signals to read the RFID tag, then sends fake data using the original RFID tag's information, thereby gaining full access to the system [21]. This attack allows the adversary to manipulate the system by injecting false data, compromising the integrity and security of the network.

\subsubsection{Routing attacks}
The adversary can manipulate routing information and distribute it across the network to create routing loops, advertise false routes, send error messages, or drop network traffic [22]. These actions can disrupt normal network operations, leading to significant inefficiencies, data loss, and compromised network integrity, effectively undermining the reliability of the entire IoT system.

\subsubsection{Sybil attack}
In a Sybil attack, a single malicious node claims the identities of multiple nodes and pretends to be those nodes. This malicious node can cause significant harm, such as distributing false routing information or disrupting the Wireless Sensor Network (WSN) election process by rigging it in its favor [23]. Such attacks can undermine the integrity of the network, leading to incorrect routing decisions, network congestion, or even a complete breakdown of network operations.

\subsubsection{Security requirements of network layer}

Although the existing core network security measures are relatively mature, certain security concerns remain particularly harmful in the context of IoT. Issues such as DoS and Distributed Denial of Service (DDoS) attacks must be effectively prevented at this layer. Additionally, communication protocols need to be highly robust to address routing attacks, congestion issues, and spoofing attacks. Ensuring the maturity and resilience of these protocols is essential to safeguarding the integrity and availability of IoT networks.

\subsection{Support Layer Security}
Support layer security operates independently from other layers, and cloud computing security is a vast and critical domain within the broader security landscape. The Cloud Security Alliance (CSA) is actively establishing standard security frameworks for cloud environments. Additionally, mechanisms like the Security Content Automation Protocol (SCAP) [24] are being developed for continuous cloud auditing, and Trusted Computing Group (TCG) [25] ensures the delivery of trusted results. Since this layer hosts IoT users' data and applications, it is crucial to protect both from security breaches. Some of the key security concerns at this layer include data breaches, unauthorized access, and vulnerabilities within the cloud infrastructure that could compromise the confidentiality, integrity, and availability of sensitive information. 

\subsubsection{Data security}

To ensure the confidentiality and security of data in the cloud, it must be safeguarded against breaches. This can be achieved by implementing tools that detect unauthorized data migration from the cloud, utilizing data loss prevention (DLP) tools, and monitoring file and database activities. Additionally, techniques such as data dispersion and data fragmentation can be employed to enhance data security in the cloud [26]. These methods help to distribute and obscure data, making it more difficult for attackers to compromise sensitive information, thereby strengthening the overall security posture of cloud-hosted IoT data and applications.

\subsubsection{Interoperability and portability}

Interoperability and portability among cloud vendors remain significant challenges today. Different vendors often use proprietary standards, which can create difficulties for users who wish to migrate their data or services from one cloud provider to another. This lack of standardization not only complicates migration but also introduces potential security vulnerabilities. The heterogeneity in cloud environments can lead to inconsistent security measures, increasing the risk of security exposures [26]. Addressing these challenges requires the development of universal standards and protocols that enable seamless interoperability and ensure consistent security across different cloud platforms.

\subsubsection{Business continuity and disaster recovery}
Cloud vendors must ensure the continuation of services during natural disasters such as floods, fires, and earthquakes. To support business continuity, the physical location of cloud data centers should be strategically chosen to minimize the impact of such calamities and to allow quick response teams easy access. Additionally, cloud providers should implement robust data backup plans to safeguard against data loss during these events [26]. These measures are essential for maintaining service availability and protecting critical business operations in the face of unexpected disasters.

\subsubsection{Cloud audit}
The Cloud Security Alliance (CSA) establishes various standards for cloud vendors to ensure robust security practices. To build and maintain user trust, continuous audits are necessary to verify that cloud vendors comply with these security standards. Regular audits help ensure that the security measures are effective and up to date, addressing any potential vulnerabilities and demonstrating a commitment to maintaining a secure cloud environment.

\subsubsection{Tenants security}

In cloud environments, multiple users' data may reside on the same physical drive, particularly in Infrastructure as a Service (IaaS) models, where different users, known as tenants, share the same physical storage. This shared physical media can pose security risks, as an adversary could potentially steal data belonging to other tenants on the same drive. Ensuring data isolation and implementing robust encryption techniques are critical to preventing unauthorized access and protecting tenants' data in such shared environments.

\subsubsection{Virtualization security}
Different cloud vendors employ various virtualization techniques, making the security of virtualization a critical concern. In some cases, communication between virtual machines (VMs) can bypass traditional network security controls, exposing the system to potential risks [26]. Additionally, secure migration of virtual machines is essential, as improper handling during migration can introduce vulnerabilities and complicate cloud audits. Ensuring that VM migration processes are secure is vital to maintaining the integrity of the cloud environment and upholding the standards required for continuous auditing and compliance.

\subsubsection{Security requirements of support layer}
In the IoT ecosystem, user data, and application instances often reside on cloud and fog nodes, making their security and privacy paramount. To protect these assets, it is crucial that cloud services adhere to established security standards, laws, and regulations set by organizations like the Cloud Security Alliance (CSA). Continuous monitoring of compliance with these standards is essential, and IoT systems should only utilize cloud services that meet CSA security requirements. Additionally, there is a need for a simple and accessible online cloud audit mechanism, enabling users to verify the security practices of their cloud vendors. This approach not only ensures compliance but also builds and maintains user trust in the cloud services they rely on.

\subsection{Application Layer Security}
Different applications within the application layer of IoT systems have varying security requirements. Currently, there is no standardized framework for constructing IoT applications, which complicates efforts to ensure consistent security across different use cases. A key characteristic of the IoT application layer is data sharing, which introduces challenges related to data privacy and access control [27]. Some of the common security issues at this layer include unauthorized access to sensitive data, inadequate access control mechanisms, and the potential for data breaches during sharing processes. Addressing these concerns requires the development of robust security protocols tailored to the specific needs of each application, ensuring both the privacy and security of the data being shared across the IoT ecosystem.

\subsubsection{Data access and authentication}

An application may have multiple users, each with different access privileges. To ensure security at the application layer, it is crucial to implement proper authentication and access control mechanisms [18]. However, this can be challenging, as discussed in Section 5, due to the varying levels of access required by different users and the need to protect sensitive data from unauthorized access. Developing and deploying robust authentication and access control systems that can accommodate these diverse requirements is essential for maintaining the security and integrity of IoT applications.

\subsubsection{Phishing attacks}
The adversary may use infected emails or malicious web links to steal legitimate user credentials and gain unauthorized access to a system using those stolen credentials [28]. This type of attack, often referred to as phishing, is a common method for compromising user accounts and can lead to significant security breaches if successful. Preventing such attacks requires vigilant user education, robust email filtering, and the implementation of multi-factor authentication to protect against unauthorized access.

\subsubsection{Malicious activeX scripts}
The adversary can send an ActiveX script to an IoT user through the internet, and if the user runs the script, it can compromise the entire system [29]. This type of attack exploits the user's trust and the script's ability to perform actions on the user's behalf, potentially leading to the execution of malicious code that can take control of IoT devices or access sensitive information. To mitigate such risks, it's important to disable or restrict the use of ActiveX controls, especially from untrusted sources, and to employ robust security measures that prevent the execution of unauthorized scripts.

\subsubsection{Malwares attack}
Attackers can target applications using malware, potentially stealing data or causing DoS attacks. Trojan horses, worms, and viruses are some of the most dangerous types of malware employed by adversaries to exploit and compromise a system [29]. These malicious programs can infiltrate IoT applications, disrupt services, and exfiltrate sensitive information, posing significant risks to the security and integrity of IoT systems. Implementing robust cybersecurity measures, including malware detection and prevention tools, is crucial to defend against these types of attacks.

\subsubsection{Security requirements of application layer}

To address security challenges at the application layer, it is essential to implement strong authentication and access control mechanisms. In addition to these measures, educating users on the importance of using strong, complex passwords [30] is also crucial. Furthermore, deploying robust antivirus software is necessary to protect against malware, including viruses, worms, and Trojan horses, which can compromise application security and integrity. These combined efforts can significantly enhance the security of IoT applications, reducing the risk of unauthorized access and malware infections.

\begin{table*}[t]
\centering
\caption{Layer Wise IoT Security and Attacks}
\begin{tabularx}{\textwidth}{|X|c|c|c|c|c|}
\hline
\textbf{Attacks} & \textbf{Perceptual Layer} & \textbf{Network Layer} & \textbf{Support Layer} & \textbf{Application Layer} & \textbf{Impact} \\ \hline
Node Tempering & \ding{51} & \ding{55} & \ding{55} & \ding{55} & High \\ \hline
Fake Node & \ding{51} & \ding{55} & \ding{55} & \ding{55} & High \\ \hline
Side Channel Attack & \ding{51} & \ding{55} & \ding{55} & \ding{55} & Medium \\ \hline
Physical damage & \ding{51} & \ding{55} & \ding{55} & \ding{55} & Medium \\ \hline
Malicious Code injection & \ding{51} & \ding{55} & \ding{55} & \ding{51} & High \\ \hline
Protecting Sensor Data & \ding{51} & \ding{55} & \ding{55} & \ding{55} & Medium \\ \hline
Mass Node authentication & \ding{51} & \ding{51} & \ding{55} & \ding{55} & High \\ \hline
Heterogeneity problem & \ding{55} & \ding{51} & \ding{51} & \ding{55} & High \\ \hline
Network Congestion problems & \ding{55} & \ding{51} & \ding{55} & \ding{55} & Medium \\ \hline
RFIDs interference & \ding{55} & \ding{51} & \ding{55} & \ding{55} & Low \\ \hline
Node jamming in WSN & \ding{55} & \ding{51} & \ding{55} & \ding{55} & Low \\ \hline
Eavesdropping Attack & \ding{55} & \ding{51} & \ding{55} & \ding{55} & Low \\ \hline
Denial of service & \ding{55} & \ding{51} & \ding{55} & \ding{55} & High \\ \hline
RFID Spoofing & \ding{55} & \ding{51} & \ding{55} & \ding{55} & High \\ \hline
Routing attacks & \ding{55} & \ding{51} & \ding{55} & \ding{55} & High \\ \hline
Sybil Attack & \ding{55} & \ding{51} & \ding{55} & \ding{55} & High \\ \hline
Data Security & \ding{55} & \ding{55} & \ding{51} & \ding{55} & High \\ \hline
Interoperability and Portability & \ding{55} & \ding{51} & \ding{51} & \ding{55} & Medium \\ \hline
Business Continuity and Disaster Recovery & \ding{55} & \ding{55} & \ding{51} & \ding{55} & Medium \\ \hline
Cloud Audit & \ding{55} & \ding{55} & \ding{51} & \ding{55} & Medium \\ \hline
Tenants Security & \ding{55} & \ding{55} & \ding{51} & \ding{55} & High \\ \hline
Virtualization Security & \ding{55} & \ding{55} & \ding{51} & \ding{55} & Medium \\ \hline
Data Access and Authentication & \ding{55} & \ding{55} & \ding{55} & \ding{51} & High \\ \hline
Phishing Attacks & \ding{55} & \ding{55} & \ding{55} & \ding{51} & Medium \\ \hline
Malicious Active X Scripts & \ding{55} & \ding{55} & \ding{55} & \ding{51} & High \\ \hline
Malware attack & \ding{55} & \ding{55} & \ding{51} & \ding{51} & High \\ \hline
\multicolumn{1}{|X|}{\textbf{Countermeasures}} & 
\multicolumn{5}{p{10cm}|}{
\begin{itemize}
  \item \textbf{Perceptual Layer:} Physical Security in nodes vicinity, Need for Lightweight Encryption Algorithms for Constrained nodes, Sensor Data privacy, Effective authentication and access control mechanisms for devices, Anti Dos attacks mechanisms.
  \item \textbf{Network Layer:} Secure Communication protocols against replay attacks, Routing attacks, Jamming attacks, Spoofing, Congestion handling, and Anti DDoS abilities in Communication protocols.
  \item \textbf{Support Layer:} Need for Continuous Cloud Audits, Implementation of Cloud Security Alliance Standards, Secure Virtualization Technologies, Tenants Separation, Storage Encryption for users data confidentiality and integrity.
  \item \textbf{Application Layer:} Secure Application Code, Educating users to use complex passwords, Access Control Mechanisms, Key Agreement, Log monitoring, File and Database monitoring tools, Anti-malware to protect applications against malware.
\end{itemize}
} \\ \hline
\end{tabularx}
\end{table*}

\section{CHALLENGES TO TRADITIONAL SECURITY SOLUTIONS IN IOT }

Security is a fundamental requirement for any user of digital media. Internet users are unlikely to share their confidential and important data on a network unless they trust its security. With the advent of cloud computing, the demand for security has increased, as users must place their trust in third-party cloud providers. To attract more users, cloud vendors must build this trust by conducting cloud audits and obtaining certifications of compliance with Cloud Security Alliance (CSA) standards or other recognized security standards.

Although traditional network security solutions are well-developed, they are not directly applicable to the IoT context due to the vast size of IoT networks, the heterogeneity of their architecture, and the resource constraints of IoT end nodes. Therefore, new security approaches specifically tailored to the unique challenges of IoT are necessary to ensure robust protection for users in this expanding digital landscape.

\subsection{Cryptographic techniques}

Currently, available cryptographic algorithms, such as symmetric key algorithms like the Advanced Encryption Standard (AES), are widely used to ensure data confidentiality and are considered highly secure. Similarly, the Rivest-Shamir-Adleman (RSA) algorithm is a frequently used asymmetric cryptographic method for digital signatures and key exchange, offering robust security. Secure Hash Algorithms (SHA) are employed to maintain data integrity, while the Diffie-Hellman (DH) protocol is used for secure key agreement. Additionally, Elliptic Curve Cryptography (ECC) is an efficient asymmetric cryptographic technique, though it is not as widely used at present [35].

While these algorithms provide strong security, they are also resource-intensive, requiring substantial CPU power and battery consumption. This makes them less feasible for use in IoT devices, which are often battery-operated and resource-constrained. Therefore, there is a pressing need to develop new cryptographic algorithms or optimize existing ones to better suit the unique requirements of IoT environments, ensuring security without compromising the efficiency and longevity of IoT devices.

\subsection{Key management}
Key management is a critical and frequently discussed research challenge in the context of cryptographic algorithms. Researchers have proposed various solutions to address this problem [36], [37], [38]. While some of these solutions are applicable to other networked systems, they are not well-suited to IoT environments due to the large scale and diverse nature of connected nodes at the device layer of the IoT architecture. As a result, key management in IoT systems remains a significant research challenge that requires further attention to develop an ideal solution that is both scalable and efficient for the unique demands of IoT networks.

\subsection{Denial of service}

Denial of Service (DoS) attacks pose a particularly severe threat in IoT environments, as they can have catastrophic consequences, such as causing loss of life if successfully launched against critical applications like smart cars [5]. Traditional DDoS detection and mitigation solutions may not be suitable for IoT systems because of the unique constraints of IoT devices, such as limited battery life and computational resources. In IoT, even allowing as few as 10 attack messages to reach sensor nodes before detecting and blocking a DoS attack could severely impact these resource-constrained devices. Existing solutions for DoS detection and mitigation [39], [40] are not fully effective in the IoT context and require further research to develop more efficient and timely methods tailored to the specific needs of IoT systems.

\subsection{Authentication and access control}

IoT heavily relies on Machine-to-Machine (M2M) communication [18]. In such communication scenarios, node authentication is crucial to ensuring security and privacy. When two or more nodes communicate for a common objective, they must first authenticate each other to prevent fake node attacks. However, the current landscape lacks an efficient authentication mechanism capable of handling the massive number of IoT devices. This gap creates a significant security vulnerability that needs to be addressed to protect IoT systems from potential attacks and ensure reliable M2M communications. Developing scalable and efficient authentication solutions for the vast number of IoT devices remains a critical research challenge that requires immediate attention.

\section{AUTHENTICATION AND ACCESS CONTROL IN IOT}

\begin{table*}[t]
\centering
\caption{Comparison of Authentication and Access Control Techniques}
\begin{tabularx}{\textwidth}{|l|X|X|X|X|}
\hline
\textbf{Ref \#} & \textbf{Authentication} & \textbf{Access Control} & \textbf{Application} & \textbf{Security} \\ \hline
Chen et al. \cite{ref6} & Nil & Group Access & Distributed IoT System & IPsec \\ \hline
Rivera et al. \cite{ref7} & OAuth 2.0 & User managed Access Model & Multi-Agent IoT system & TLS \\ \hline
Ouaddah et al. \cite{ref9} & Nil & Organization bases access control & Inter Organizations & Web services \\ \hline
Gaikwad et al. \cite{ref10} & Kerberos & Nil & Smart Homes (IoT) & AES, SHA1 \\ \hline
Periera et al. \cite{ref11} & Credentials, shared key, password & Service Level Access control & Nil & DTLS light implementation \\ \hline
Mahalle et al. \cite{ref12} & Group Authentication & Nil & Wi-Fi & Lightweight cryptographic function \\ \hline
Panwar et al. \cite{ref13} & Digital certificates & Nil & Nil & DTLS \\ \hline
Santoso et al. \cite{ref31} & Elliptic Curve Cryptography & Nil & Smart Homes (IoT) & Encryption, Wi-Fi \\ \hline
Lee et al. \cite{ref32} & Lightweight Cryptography & Nil & Nil & Lightweight XOR operation \\ \hline
Park et al. \cite{ref33} & Simple certificates & Nil & Nil & PKI \\ \hline
Zhao et al. \cite{ref34} & Elliptic Curve Cryptography & Nil & Nil & SHA1 \\ \hline
\end{tabularx}
\end{table*}

The security of the IoT is a highly active area of research today, with a vast number of publications highlighting various security and privacy issues within IoT systems. Due to the enormous number of IoT devices and the machine-to-machine (M2M) communication model that IoT relies on, traditional authentication and authorization techniques are not suitable. In IoT environments, devices must authenticate each other before exchanging any information, which is a significant challenge given the massive scale of IoT networks.

Several researchers have proposed solutions related to device authentication and access control in IoT. For instance, Chen et al. [6] proposed a Capability-based access control model for distributed IoT environments. This model supports group access by using a single token and guarantees end-to-end security through IPsec. A requester can use a single token to communicate with any device in a group that offers common services. Each device within the group is identified by a unique local identifier (ULA), and the network prefix of the access group is used as the access group identifier. The devices in the group can verify the token using their ULA and the prefix in the token, and access control can also be provided based on the requester's ULA within the token.

Existing standards such as TLS and PKI have addressed the first three domains of security, namely confidentiality, integrity, and authentication. However, access control requires further attention. For example, in multi-agent systems, different agents have different roles and therefore require different access levels. Rivera et al. [7] proposed the use of the User-Managed Access model, a profile of OAuth 2.0, which provides different access levels to various agents based on their roles.

OUADDAH et al. [9] proposed a novel access control framework for IoT environments called "SmartOrBAC," which is based on the OrBAC model. This model uses web services (RESTful approach) to enforce security policies. Organization-based Access Control (OrBAC) has some limitations, such as better performance in centralized systems, lack of support for collaboration between organizations and sub-organizations, and inability to translate security policies into access control mechanisms. To address these limitations, SmartOrBAC, an extension of OrBAC, was proposed. SmartOrBAC uses web services to ensure secure collaboration between different organizations and emphasizes using RESTful APIs for exchanges between organizations, as it is a lightweight mechanism.

The interactions between organizations are defined by agreements between them, with access rules being established according to the OrBAC format. In SmartOrBAC, these contracts are not predefined but can be established dynamically and spontaneously. SmartOrBAC provides efficient access control for collaborative entities, especially in low-power and energy-constrained scenarios, such as those found in IoT.

Gaikwad et al. [10] implemented a three-level secure Kerberos authentication system for smart home IoT applications, utilizing the Secure Hash Algorithm (SHA-1) and Advanced Encryption Standard (AES) for security. However, Kerberos is not considered a sustainable solution for authentication in IoT, and AES is impractical for resource-constrained IoT devices due to its high computational demands.

Periera et al. [11] proposed a service-level access control framework specifically designed for power-constrained IoT devices. This framework allows fine-grained access control on a per-service basis by merging concepts from Kerberos and RADIUS access control systems. It integrates the best features of Kerberos, Constrained Application Protocols (CoAP), and RADIUS to create a low-power platform for access control and authentication. The framework operates in two steps: first, the user is authenticated using credentials such as a shared key, password, or other validators. Upon successful authentication, the CoAP-NAS (Network Access Server) is informed about the user's permissions, ticket timeout, group membership, etc. The CoAP-NAS then issues a ticket to the user for future requests. In the access control step, the server will only respond if the request message contains a valid ticket; otherwise, it will generate an error message.

Mahalle et al. [12] introduced a lightweight, secure, and scalable Threshold Cryptography-based Group Authentication (TCGA) scheme for IoT, which verifies the identity of all nodes in group communication. This group authentication reduces the overhead of handshakes, ensuring lower resource usage and helping to conserve power. The scheme is also secure against man-in-the-middle attacks.

Panwar et al. [13] proposed a security mechanism for IoT that employs digital certificates with Datagram Transport Layer Security (DTLS). In this approach, authentication is conducted through digital certificates provided by a certificate authority, which enhances security and replaces the pre-shared key mechanism traditionally used in DTLS. The client/server authentication process involves several steps:
1. The client sends a request to the server.
2. The server responds by sending its certificate to the client.
3. The client verifies the certificate by decrypting it with the server’s public key.
4. After successful verification, the client sends its own certificate to the server.
5. The server then verifies the client’s certificate using the same procedure, allowing secure communication to commence.

Santoso et al. [31] proposed a robust security scheme for smart home systems, based on the AllJoyn framework and utilizing Elliptic Curve Cryptography (ECC) for the authentication process. The system operates over a Wi-Fi network and includes a Wi-Fi gateway node responsible for initial system configuration, device authentication, and providing a means for users to control the system via an Android mobile application. The authentication process consists of two steps: first, the mobile device authenticates with the IoT device (where the user loads the identity and pre-shared key, and after mutual authentication, home credentials are shared with the IoT device). Second, the IoT device connects to the gateway, which authenticates the device using information provided by the mobile device. Once authentication is complete, encrypted communication between the devices takes place.

Lee et al. [32] introduced a lightweight authentication protocol that enhances the security of the original RFID system for IoT applications. The existing RFID protocol had a significant security flaw as it performed authentication without encryption. To address this, Lee et al. proposed a lightweight cryptographic protocol based on the XOR method, where encrypted passwords are used for authentication, significantly improving the security of the system.

Park et al. [33] proposed a framework that simplifies the certificate structure for IoT device authentication by creating a permit code structure, which is more suitable for small, resource-constrained IoT devices. Traditional certificates are based on digital signatures, which are difficult to implement on these devices. The permit code approach offers a more manageable solution in IoT environments.

Zhao et al. [34] developed an asymmetric mutual authentication scheme for IoT, where authentication occurs between the terminal node and the platform. The proposed scheme combines SHA-1 and feature extraction techniques to enhance IoT security while reducing both computational and communication costs. This approach not only strengthens the security of IoT systems but also increases their efficiency, making it a viable solution for resource-constrained devices.

\section{Conclusion}

The security of the Internet of Things is a highly active research area in both industry and academia, demanding further exploration to address the various security challenges within IoT systems. This paper investigates the primary security issues across each layer of the four-layer IoT architecture: perceptual layer, network layer, support layer, and application layer. Notably, the security concerns in the support layer have not been thoroughly explored in the context of IoT, and this paper provides a comprehensive study of these issues. Additionally, we propose brief countermeasures to address different security challenges and secure IoT systems.

We also discuss the limitations of legacy security solutions when applied to IoT, particularly regarding authentication and access control mechanisms. Traditional authentication methods are inadequate for IoT due to the resource-constrained nature and massive scale of IoT devices. As a result, new authentication mechanisms are needed to effectively authenticate these constrained devices in machine-to-machine (M2M) communications. This paper presents a study of the state-of-the-art authentication and access control mechanisms specifically designed for IoT. This comprehensive study aims to guide researchers in identifying areas where efforts should be focused to develop robust security solutions for IoT.


\end{document}